\documentclass[aps,pra,twocolumn,superscriptaddress,epsfig]{revtex4}

\usepackage{graphicx}
\usepackage{amsmath}
\usepackage{epsfig,times}
\usepackage{dcolumn}
\usepackage{bm}
\usepackage{color}
\usepackage{epstopdf}
\usepackage{amssymb}
\usepackage{amstext}
\usepackage{latexsym}
\usepackage{hyperref}
\usepackage{amsfonts}
\usepackage{psfrag}

\usepackage{mathptmx} 

\newcommand{\beq}{\begin{equation}}
\newcommand{\eeq}{\end{equation}}
\newcommand{\bea}{\begin{eqnarray}}
\newcommand{\eea}{\end{eqnarray}}

	\newcommand{\tr}[1]{\textrm{Tr} \left[ {#1} \right]} 

	\newcommand{\e}[1]{e^{ {#1}}} 

	\newcommand{\be}{\begin{equation}}
	\newcommand{\ee}{\end{equation}}

	\newcommand{\ket}[1]{{\left\vert {#1} \right\rangle}}	
	\newcommand{\bra}[1]{{\left\langle {#1} \right\vert}}

\usepackage{amsthm}
\usepackage{mathrsfs}
\usepackage{lipsum}

\def\ln{{\rm ln}}

\def\la{\langle}
\def\ra{\rangle}
\def\om{\omega}

\def\tr{\rm{Tr}}

\newcommand{\beqa}{\begin{eqnarray}}
\newcommand{\eeqa}{\end{eqnarray}}

\begin{document}

\title{
More bang for your buck: 
Towards super-adiabatic quantum engines}
\author{A. del Campo}
\affiliation{Theoretical Division,  Los Alamos National Laboratory, Los Alamos, NM 87545, USA}
\affiliation{Center for Nonlinear Studies,  Los Alamos National Laboratory, Los Alamos, NM 87545, USA}
\author{J.~Goold}
\affiliation{Clarendon Laboratory, Department of Physics, University of Oxford, Oxford OX1 3PU, United Kingdom}
\author{M. Paternostro}
\affiliation{Centre for Theoretical Atomic, Molecular and Optical Physics, School of Mathematics and Physics, Queen's University Belfast, BT7 1NN Belfast, United Kingdom}
\begin{abstract}
The reversible nature of thermodynamical cycles is an idealisation based on the assumption of perfect quasi-static dynamics. As a consequence of this assumption, ideal engines operate at the maximum efficiency but have zero power. Realistic engines, on the other hand, operate in finite-time and are intrinsically irreversible, implying {\it friction} effects at short cycle times. The engineering goal is to find the maximum efficiency allowed at the maximum possible power \cite{Calborn,Esposito}. In the current technological age, with our ability to manipulate devices at the nanoscale and beyond, one must understand the consequences of engines which operate at the quantum mechanical level. In this domain one cannot avoid the emergence of both quantum and thermal fluctuations \cite{jrev,campisi} which drastically alter the energetics of the cycle \cite{sekimoto}. Very recently, it has been shown that the Hamiltonian of a quantum system may be manipulated in such a way as to mimic an adiabatic process via a non-adiabatic shortcut~\cite{DR03,Berry09,Chen10,STAreview}. 
A surge of experimental progress has demonstrated several of these proposals in the laboratory \cite{Schaff1,Schaff2,hfqd}. In this paper we show that, by utilising shortcuts to adiabaticity~\cite{STAreview} in a quantum engine cycle, one can engineer a thermodynamical cycle working at finite power and zero friction. Our findings are elucidated using a harmonic oscillator undergoing a quantum Otto cycle.

\end{abstract}
\maketitle

Thermodynamics is the study of heat and its interconversion to mechanical work. It successfully describes the ``equilibrium'' properties of macroscopic systems ranging from refrigerators to black holes~\cite{refrig}. The technological revolution potentially embodied by the incorporation of information technology is motivating the consideration and realisation of quantum devices going all the way down to the micro- and nano-scale~\cite{nanotech}. This has forced us to revise our interpretation of thermodynamics so as to include {\it ab initio} both quantum and thermal fluctuations, which become prominent at those working scales. In fact, far from equilibrium, such fluctuations become dominant and cannot be neglected. In turn, thermodynamical quantities such as work and heat become inherently stochastic and should be reformulated as such. 

Recently discovered work fluctuation theorems (FTs) and the corresponding framework, which is known to hold both in quantum and classical systems, are extremely useful for the task of setting up a quantum apparatus for thermodynamics~\cite{tasaki}. FTs demonstrate that information on the equilibrium state of the systems is encoded in the associated probability distributions, and can be interpreted as refined statements of the second law applicable at the micro- and nano-scale helping us understand the origin of microscopic irreversibility in systems undergoing finite-time transformations~\cite{jrev,campisi}. Here we address the important related question of whether or not FTs can help us shed light on limitations or possible advantages of a quantum device operating in finite-time. A convenient platform to look for a quantitative answer is provided by quantum engine cycles, for which the thermodynamic laws must be recast appropriately~\cite{mahlerbook,scovil,alicki,ion_engine}. In fact, although its working principles might well be quantum mechanical, the efficiency of a reversible engine would always be limited by the second law, which makes the quantum version of cycles remarkably similar to their classical counterparts.

The assumed reversibility (quasi-stationarity) of an engine cycle, which implies its zero-power nature due to its infinitely long cycle-time, crashes with the reality of any practical machine, either quantum or classical. This is due to the finite-time operation which exposes it to the effects of friction-induced losses. While Ref.~\cite{kosloff_lube} proposes the use of systematic noise to suppress frictional losses in the expansion and compression stages of a quantum Otto cycle, here we devise an innovative way to run a finite-time (and power) quantum cycle based on the use of shortcuts to adiabaticity~\cite{DR03,Berry09,Chen10,STAreview}. As we will show, such Hamiltonian-engineering techniques, which have recently found considerable interest in the quantum community, allow us to drive the expansion and compression stages of a cycle, which are prone to frictional effects, virtually without any loss affecting the performance of a quantum engine within the finite-time of its cycle. Drawing inspiration from a recent ion-trap proposal \cite{ion_engine}, we will provide an example of a finite-time, fully frictionless quantum Otto cycle where the working medium is a quantum harmonic oscillator.

\noindent
\section{Quantum Otto Cycle} 
\begin{figure}[!t]
\includegraphics[width=0.9\columnwidth]{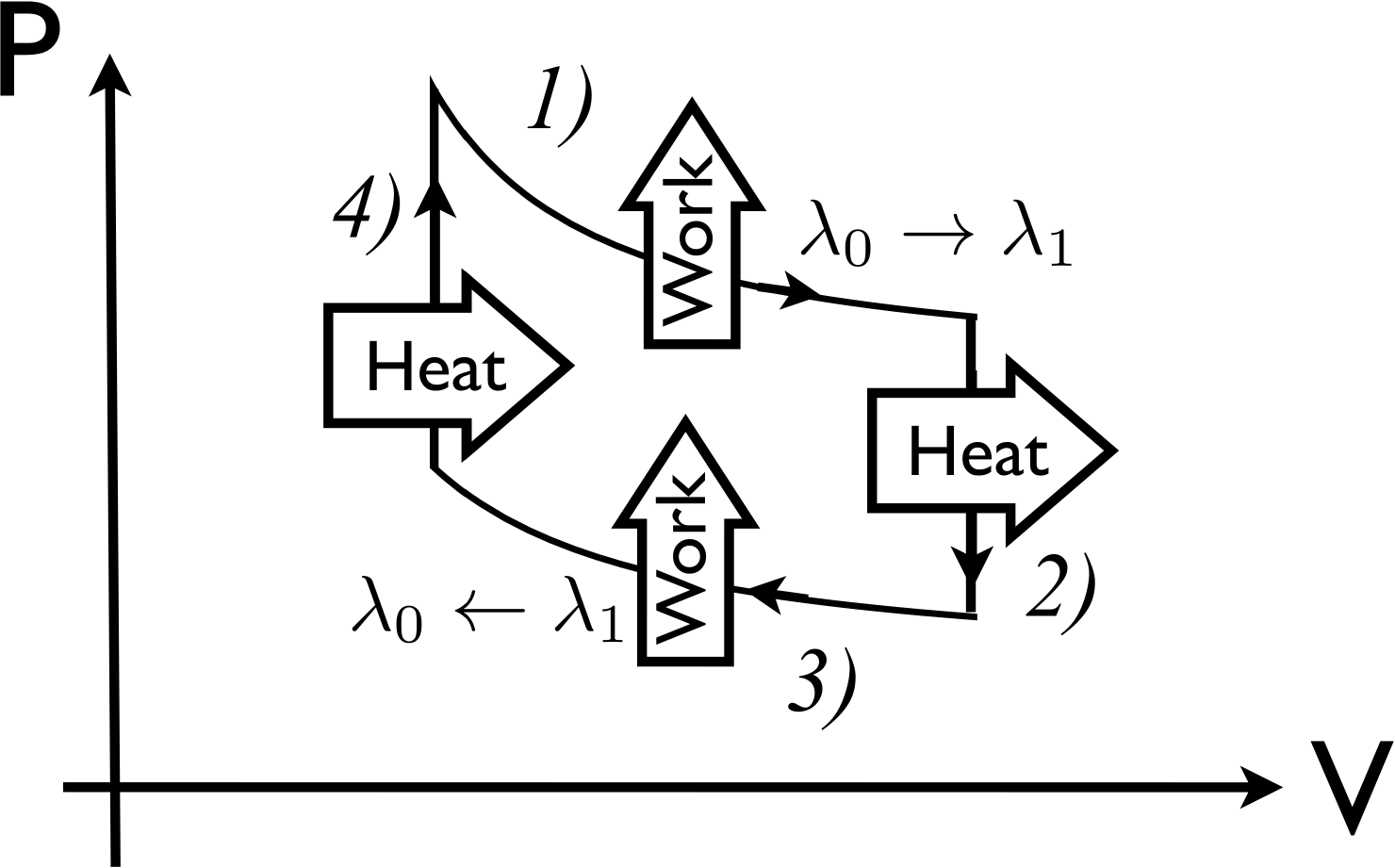}
\caption{{Pressure-volume diagram of a (quantum) Otto cycle}. The numbers relate the processes to the description of each step given in the main text. We identify the steps where heat enters (exits) the working medium and those where work is performed by (done onto) it as a result of a corresponding change in the work parameter $\lambda(t)$.}
\label{Otto}
\end{figure}

In an Otto engine, a working medium (coupled alternatively to two baths at different temperatures $T_{i}, i=1,2$) undergoes a four-stroke cycle. In its quantum version, the state of the working medium is described by a density operator $\rho(\lambda(t))$ that is changed by the Hamiltonian $\hat{\cal H}(\lambda(t))$. Here, $\lambda(t)$ is an adjustable {\it work parameter}, typical of the specific setting used to physically implement $\hat{\cal H}(\lambda(t))$, whose value determines the equilibrium configuration of the system. As illustrated in Fig.~\ref{Otto}, the cycle steps are as follows:\\
{\it 1)} {\it An adiabatic expansion} performed by the change of the work parameter $\lambda_{0}\equiv\lambda(0)\to\lambda_1\equiv\lambda(\tau_1)$, where $\tau_1$ is the time at which this step ends. As a result of this transformation, work is extracted from the medium due to the change in its internal energy.\\ 
{\it 2)} {\it A cold isochore} where heat is transferred from the working medium to the cold bath. This is associated with a heat flow from the medium to the cold reservoir it is in contact with.\\
{\it 3)} {\it An adiabatic compression} performed by the reverse change of the work parameter $\lambda_{1}\rightarrow\lambda_{0}$ and during which work is done on the medium.\\
{\it 4)} {\it A hot isochore} during which heat is taken from the hot reservoir by the working medium. 

\noindent
If the engine is run in a finite-time, {\it i.e.} if we abandon the usual quasi-static assumption, friction is inevitably generated along the expansion and compression steps. We will elucidate the nature of this friction later on. In addition, one may (realistically) assume imperfect heat conduction during the isochores. Under such conditions, the work $W$ done by/on the engine across the cycle, and the heat $Q$ exchanged by the working medium with the baths, become stochastic quantities with associated probability distributions $P(K;t)~(K=W,Q)$. The efficiency of the engine is then defined as the ratio between the  average total work 
per cycle and the average heat received from the hot bath, according to the equation
\beq
\label{efficiency}
{\cal E}=-\frac{\langle W\rangle_{1}+\langle W\rangle_{3}}{\langle Q\rangle_{4}},
\eeq
where $\langle K\rangle_j$ is the average value of the quantity $K=Q,W$ during step $j=1,\dots,4$ of the cycle. 
The power of the engine is defined accordingly as 
\beq
{\cal P}=-\frac{\langle W\rangle_{1}+\langle W\rangle_{3}}{\sum^4_{j=1}\tau_{j}},
\eeq
where $\tau_{j}$ is the time needed by the $j^{\rm th}$ step. In what follows we consider the case where friction only occurs along the adiabatic transformations. In particular, we  neglect fluctuations in the heat flow, so that $\langle Q\rangle\simeq Q$, and we assume that the thermalisation process associated with the 
isochoric transformations occur in a timescale that is much quicker than any other one in the dynamics of the cycle. Under these assumptions, the power of the engine can be approximately written as
\beq
{\cal P}\simeq-\frac{\langle W\rangle_{1}+\langle W\rangle_{3}}{\tau_{1}+\tau_{3}}.
\eeq

\noindent 
\section{Finite-time thermodynamics} Before we quantify the efficiency of the engine, it is necessary to properly define the probability distribution of work of which $\langle W\rangle$ is the first moment. We consider a dynamical system described by $\hat{\cal H}(\lambda(t))$. The system is prepared by allowing it to equilibrate with a heat reservoir at inverse temperature $\beta$ with $\lambda(t\leq0)=\lambda_0$. The initial state of the system $\rho_0$ is the Gibbs state
\begin{equation}
\rho_\textrm{G}(\lambda)=\frac{\e{-\beta \hat{\cal H}(\lambda)}}{{Z}(\lambda)}
\nonumber
\end{equation}
with ${Z}(\lambda):=\tr[{\e{-\beta \hat{\cal H}(\lambda)}}]$ the associated partition function. At $t=0$, the system-reservoir coupling is removed and a protocol is performed on the system taking the work parameter from its initial value $\lambda_0$ to $\lambda_1$ at a later time $t=\tau$. For instance,  the process could be  either the expansion or the compression step of the Otto cycle, 
represented by a change of the parameter $\lambda$ of the system Hamiltonian $\hat{\cal H}(\lambda)=\sum_n \varepsilon_n(\lambda) \ket{n(\lambda)}\bra{n(\lambda)}$, where $\ket{n(\lambda)}$ is the $n^\textrm{th}$  eigenstate with eigenvalue $\epsilon_n(\lambda)$.
As discussed in Ref.~\cite{lutz}, in this context the very concept of work should be reformulated in a way to account for both the statistics of the initial state of the system and the inherently non-deterministic nature of quantum mechanics and measurements. 
The explicit expression for the work distribution function $P(W;t)$ reads \cite{lutz}
\beqa
\label{P(W)}
P(W;t)=\sum_{k,n}\delta[W-(\varepsilon_k(\tau)-\varepsilon_n(0))]p_{nk}^tp_n^0.
\eeqa
Here, letting $|n(t)\ra=|n(\lambda(t))\ra$, $p_{nk}^t=|\la k(t)|\hat{U}(t,0)|n(0)\ra|^2$ is the  transition probability $|n(0)\ra \rightarrow |k(t)\ra$ under the evolution operator $\hat{U}(t,0)$ associated with $\hat{\mathcal{H}}(t)$,
and $p_n^0$ is the occupation probability of the initial mode $|n(0)\ra$ which for a Gibbs ensemble reads $p_n^0=e^{-\beta\varepsilon_n(0)}/\sum_n e^{-\beta\varepsilon_n(0)}$. 
The $m^{\rm th}$ moment of the distribution $P(W;t)$ is $\la W^m\ra=\sum_{k,n}[\varepsilon_k(t)-\varepsilon_n(0)]^mp_{nk}^tp_n$. The average work extracted or done up to time $t$ is found for $m=1$.

For finite systems, the statistical nature of work requires the second law of thermodynamics to be revised to the form
$\langle W \rangle\ge\Delta{F}$, with $\Delta F$ the change in free energy and the equality holding for a quasi-static isothermal process (the inequality holding strictly for all quasi-static processes performed without the coupling to a thermal reservoir). For all non-ideal processes, the deficit between the average work $\langle W \rangle$ and the variation in free energy can be accounted for by the introduction of the average irreversible work $\langle W_{\rm irr}\rangle$ as $\langle W \rangle = \langle W_{\rm irr} \rangle+\Delta F$. We anticipate that the behavior of $\la W\ra$ will be contrasted, later on in this paper, with the average work $\la W_{\rm ad}\ra$ performed onto or made by the system when a shortcut to adiabaticity is implemented. For such quantity, $\la W_{\rm ad}\ra>\Delta F$, in the absence of a heat bath.
For a closed quantum system, the incoming heat flow is null and the entropy change due to the irreversibility of the process $\Delta S_{\rm irr}$
is 
\begin{equation}
\Delta S_\textrm{irr}=\beta(\langle W\rangle-\Delta F)=\beta\langle W_\textrm{irr} \rangle,
\label{eq:irrS}
\end{equation}
which can be recast as $\Delta S_{\rm irr}=S(\rho_t||\rho_t^{\rm eq})$~\cite{lutz2} with $S(\rho_A||\rho_B)=\tr(\rho_A\ln\rho_A-\rho_A\ln\rho_{B})$ the relative entropy between two density matrices $\rho_{A}$ and $\rho_B$~\cite{vedral}, $\rho_t$ the time-evolving state, and $\rho_t^{\rm eq}=e^{-\beta \hat{\mathcal{H}}(t)}/{\rm Tr}[e^{-\beta \hat{\mathcal{H}}(t)}]$ the corresponding equilibrium reference state at the initial temperature $1/\beta$. 
Here, $\langle W_{\rm irr} \rangle$ quantifies the degree of friction caused by the finite-time protocol on the expansion or compression stage of the engine cycle at hand. When a bath is reconnected this friction is manifested by dissipation into the bath and hence the decrease in the overall efficiency of the motor. For simplicity and for the point of demonstration we allow only this form of irreversibility in our engine cycle although in principle the same analysis can be done for fluctuating heat flows \cite{hexchange1,hexchange2}.

\noindent 
\section{Friction-free finite-time engine} Recently there has been a significant amount of work devoted to the design of so-called super-adiabatic protocols, {\it i.e.} shortcuts to states which are usually reached by slow adiabatic processes~\cite{DR03,Berry09,STAreview}. 
A typical approach for shortcuts to adiabaticity is to use {\it ad hoc} dynamical invariants to engineer a Hamiltonian model that connects a specific eigenstate of a model from an initial to a final configuration determined by a dynamical process. Here we will rely on an approach based on engineered non-adiabatic dynamics achieved using self-similar transformations \cite{Chen10,delcampo11}.

Let us consider a quantum harmonic oscillator with time-dependent frequency $\omega(t)$  as the working medium of the engine cycle~\cite{Chen10}.
The Hamiltonian model that we consider is thus $\hat{\cal H}(t)=\hat{\cal H}[\om(t)]={\hat{p}^2}/({2m})+{m\omega^{2}(t)\hat{x}^2}/{2}$, where $\hat x$ and $\hat p$ are the position and momentum operators of an oscillator of mass $m$. Inspired by the scheme put forward in~\cite{ion_engine}, we will use the tunable harmonic frequency to implement the compression and expansion steps of the Otto cycle. In line with the experimental proposal for the realisation of a microscopic Otto motor put forward in~\cite{ion_engine}, the frequency of the harmonic trap embodies the volume of the  chamber into which the working medium is placed, while the corresponding pressure is defined in terms of the change of energy per unit frequency.

Needless to say, in the compression or expansion stage of the Otto cycle, the frequency of the trap will have to be varied, so that $\omega(t)$ takes here the role of the work parameter $\lambda(t)$ introduced when discussing FTs. We now suppose to subject the working medium to a change in the work parameter occurring in a time $\tau$ and corresponding to, say, one of the friction-prone steps of the Otto cycle. Our goal is to design an appropriate shortcut to adiabaticity to arrange for a fast, frictionless evolution between the equilibrium configuration of the working medium at $t=0$ and that at $t=\tau$. In order to do this, we remind that the wavefunction $\phi_n(x,t=0)=\langle{x}|n(0)\rangle$ of an initial eigenstate $\ket{n(0)}$ of $\hat{\cal H}(0)$ is known to follow the self-similar evolution~\cite{Chen10}  
\begin{equation} 
\label{scaling}
	\phi_n(x,t)=\frac{1}{\sqrt{b(t)}}\exp\left(\! i\frac{m\dot{b}(t)x^2}{2\hbar b(t)}-i\frac{\varepsilon_n(0)\eta(t)}{\hbar}\!\right)\la x/b(t)|n(0)\ra,
\end{equation}
where $\eta(t)=\int_{0}^tdt'/b^2(t')$, $\varepsilon_n(0)$ is the energy of the eigenstate being considered at $t=0$, and the scaling factor $b$ is the solution of the Ermakov equation  
\begin{equation}
\label{ermakov}
\ddot{b}(t)+\omega^{2}(t)b^{2}(t)={\omega^{2}_{0}}/{b^{3}},
\end{equation}
with the initial conditions $b(0)=1$ and $\dot{b}(0)=0$. The shortcut to adiabaticity that we seek is then found by inverting the Ermakov equation and complementing the previous set of boundary conditions with $\dot{b}(0)=\ddot{b}(0)=\dot{b}(\tau)=\ddot{b}(\tau)=0$, and $b(\tau)=\sqrt{\omega_0/\omega_f}$ with $\omega_{0}=\omega(0)$ and $\omega_f=\omega(\tau)$. Instances of solutions to this problem can be found as illustrated in the Appendix, where we give the explicit form of the scaling factor $b(t)$ such that the finite-time dynamics that takes the initial state $\phi_n(x,t=0)=\la x|n(0)\ra$ to the final one $\phi_n(x,t=\tau)=\la x|n(t)\ra=\la x/b(\tau)|n(t=0)\ra/\sqrt{b(\tau)}$ actually mimics the wanted adiabatic evolution (albeit for any $t\in(0,\tau)$, $\phi_n(x,t)$ is in general different from the eigenstate $\ket{n(t)}$ of $\hat{\cal H}(t)$).
The choice of a harmonic oscillator is not a unique example as similar self-similar dynamics can be induced in a large family of many-body systems \cite{delcampo11} and other trapping potentials, such as a quantum piston \cite{DB12}.

\begin{figure}[t]
\centering{\includegraphics[width=0.95\linewidth]{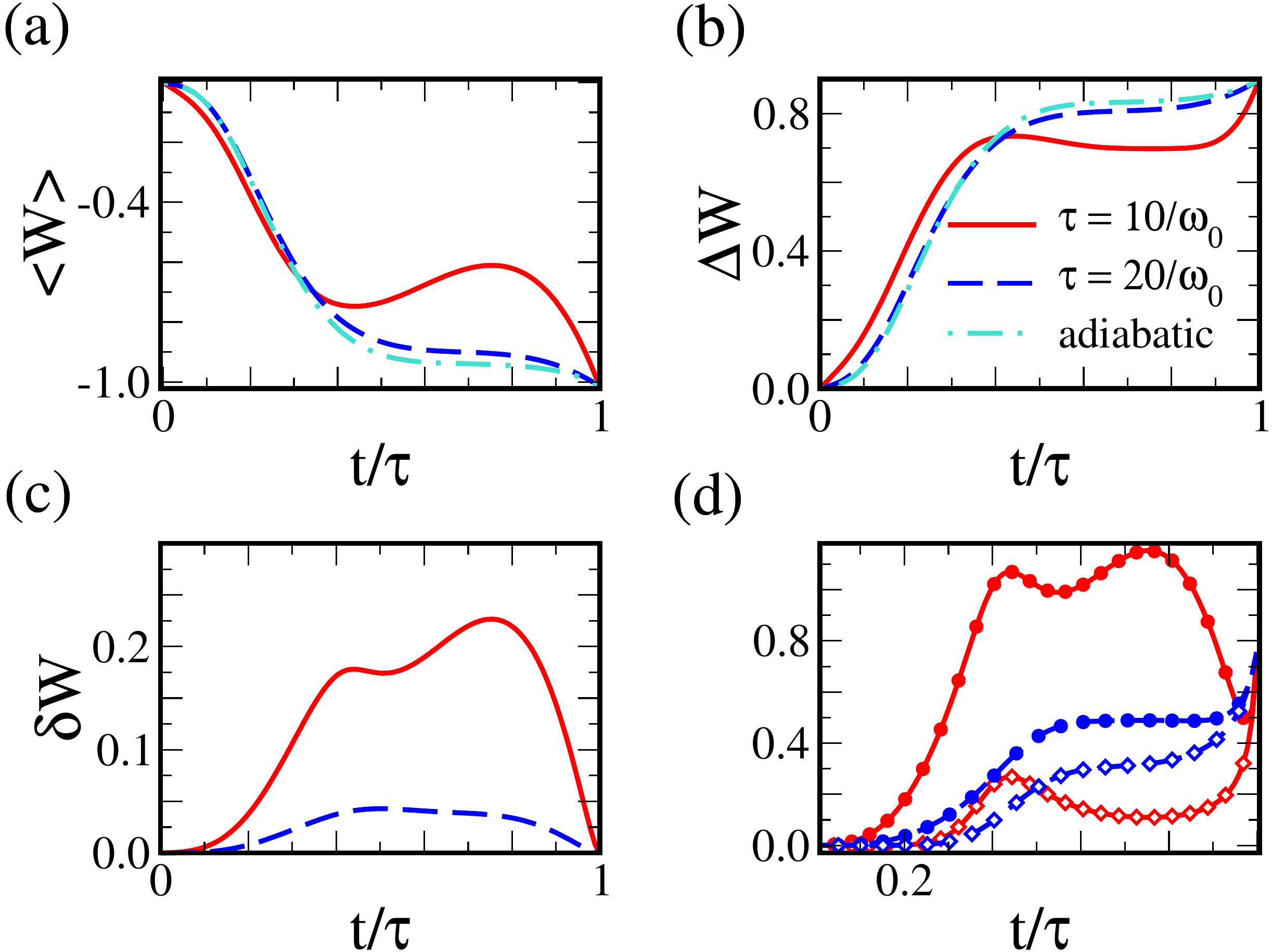}}
\caption{
{Work fluctuations along a shortcuts to an adiabaticity expansion.}
(a) Average work; (b) Standard deviation of the work;  (c) Nonequilibrium deviations from the adiabatic average mean work; (d) We show $S(\rho_t||\rho_t^{\rm eq})/\beta$ ($\bullet$) and $S(\rho_t^{\rm ad}||\rho_t^{\rm eq})/\beta$ ($\diamond$)  [cf. Eq.~\eqref{dwsrel}] for the same processes shown in the other panels.
All quantities are plotted in units of $\hbar\om_0$ ($\beta=1$). 
}
\label{eradue}
\end{figure}

Let us consider the fluctuations induced in the expansion and compression stages of the Otto cycle when the above shortcut to adiabaticity is 
implemented. Let us consider a driving  Hamiltonian 
with instantaneous eigenstates  $|n(t)\ra$ and  eigenvalues $\varepsilon_n(t)$. 
In the adiabatic limit, the corresponding transition probabilities $p_{nk}^t$ tend to $|\la n(t)|k(t)\ra|^2=\delta_{k,n}(t)$ for all $t\in[0,\tau]$. The average work simplifies then to $\la W_{\rm ad}(t)\ra=\sum_{n}[\varepsilon_n(t)-\varepsilon_n(0)]p_n=\frac{\hbar[\om(t)-\om_0]}{2}{\rm coth}\frac{\beta \hbar\om_0}{2}$. On the other hand, in a shortcut to adiabaticity, only the weaker condition 
$p_{nk}^t=|\la \phi_n(t)|k(t)\ra|^2\rightarrow\delta_{k,n}(t\rightarrow 0,\tau)$ holds.
For the time-dependent harmonic oscillator, it follows that
\beqa
 \la W\ra=\frac{\hbar}{2}\left[\!\frac{\dot{b}^2(t)\!+\!\om^2(t)b^2(t)\!+\!{\om_0^2}/{b^2(t)}}{2\om_0}-{\om_0}\right]{\rm coth}\frac{\beta \hbar\om_0}{2}.
\eeqa
In the adiabatic limit $\dot{b}(t)\rightarrow 0$ and $b(t)\rightarrow b_{\rm ad}(t)=[\om_0^2/\om^2(t)]^{1/4}$. 

Figure 2(a) shows that the average work $\la W\ra$ along a shortcut to an adiabatic expansion in comparison with in the corresponding adiabatic processes $\la W_{\rm ad}(t)\rangle$ (the  behaviour observed during a shortcut to a compression is mirrored in time). 
It is very important to stress that $\langle W\rangle$ is the work done on either adiabat until the reconnection with the bath, i.e. just prior to the isochoric heating or cooling stage. The standard deviation of the work distribution  $\Delta W=[\la W^2\ra-\la W\ra^2]^{1/2}$ is displayed in Fig. 2(b). In turn, this provides a further characterisation of the work fluctuations along the shortcut through the width of $P(W;t)$. 
It is interesting to notice that upon completion of the stroke,  the non-equilibrium deviation of both the average work and the standard deviation from the adiabatic trajectory disappear.  

We shall now analyse the non-equilibrium deviation $\delta W=\la W\ra-\la W_{\rm ad}(t)\ra$ with respect to the adiabatic work $\la W_{\rm ad}(t)\ra$. 
Note that this expression is equivalent  to the  deviation of the mean energy of the motor  along the super-adiabats from its (instantaneous) adiabatic expression. 
For an isothermal reversible process $\la W_{\rm ad}\ra=\Delta F$ and 
$\delta W=\la W_{\rm irr}\ra$.
Differently, for the adiabatic dynamics associated to stages 1 and 3 of the Otto cycle, 
conservation of the population in $|n(t)\ra$ is satisfied provided that $\beta_t=\beta_0\varepsilon_n(0)/\varepsilon_n(t)$, as it is the case for  a large-class of self-similar processes (here, $\beta_t$ is introduced by noticing that the physical adiabatic state at time $t$ is characterised by the occupation probabilities $p^t_n=e^{-\beta_t\varepsilon^t_n}/\sum_ne^{-\beta_t\varepsilon^t_n}$)~\cite{Chen10,delcampo11,DB12}. As a result, the reference state $\rho_t^{\rm eq}$ is not the physical instantaneous equilibrium state $\rho_t^{\rm ad}=\sum_np_n^0|n(t)\ra\la n(t)|$ resulting from the adiabatic dynamics, and we find
\be
\label{dwsrel}
\delta W=\frac{1}{\beta}[S(\rho_{t}||\rho_{t}^{\rm eq})-S(\rho_{t}^{\rm ad}||\rho_{t}^{\rm eq})].
\ee
%
%
From this result, it is clear that, in general, $\delta W\neq{0}$. 
However, it is straightforward to check that, at the final time of the process $t=\tau$, we have $p_{nk}^{\tau}=\delta_{k,n}$, which implies $\delta W=0$ and, in turn, the frictionless nature of the process [cf. Fig.~\ref{eradue}(c)].  
The time-evolution of the different contribution to $\delta W$, {\it i.e.} $S(\rho_{t}||\rho_{t}^{\rm eq})/\beta$ and $S(\rho_{t}^{\rm ad}||\rho_{t}^{\rm eq})/\beta$, are displayed in Figure 2(d).
This result is remarkable in the context of the quantum Otto cycle: If the baths are reconnected at just the right time $\tau$ after both the compression and expansion stages, then the efficiency of an ideal reversible engine can be reached in finite-time, therefore implementing a perfectly frictionless finite-time cycle. As we have built our engine so that friction is the only source of irreversibility, the super-adiabatic engine clearly reaches the maximum efficiency of an ideal quasi-static engine in a finite-time only. 

Let us address a final important point. The efficiency in Eq. \eqref{efficiency} of an Otto cycle 
  diminishes explicitly with the breakdown of adiabaticity \cite{ion_engine}. In contrast, the super-adiabatic engine put forward in this proposal does achieve the maximum possible value
${\cal E}=1-{\om(\tau)}/{\om(0)}$.
It should be noted, quite strikingly, that if unlimited resources are available, there is no fundamental lower-bound on the running time of the adiabats $\tau_{1,3}$.  However, it is worth taking a pragmatic approach here and attempt at the quantification of the energy costs associated with the running of our super-adiabatic engine. To this end, we have considered the time-averaged  dissipated work $\la\delta W\ra=\tau^{-1}\int_0^{\tau}\delta W dt$ for $\tau>\tau_c$, ensuring $\om^2(t)>0$ for all $t\in[0,\tau]$. The cut-off time $\tau_c$ was taken to be the maximum running time along the shortcut of the super-adiabat before the trap is inverted. Indeed, when this occurs, the adiabatic eigenenergies are not well defined, implying the break-down of our formalism.  
The explicit expression for $\la\delta W\ra$  are reported in the Appendix. 
Figure \ref{dissW} shows that the cost of running the super-adiabatic engine exhibits a neat power-law behaviour $\la\delta W\ra\sim 1/\tau$ for a wide range of parameters.
\begin{figure}[t]
\epsfig{file=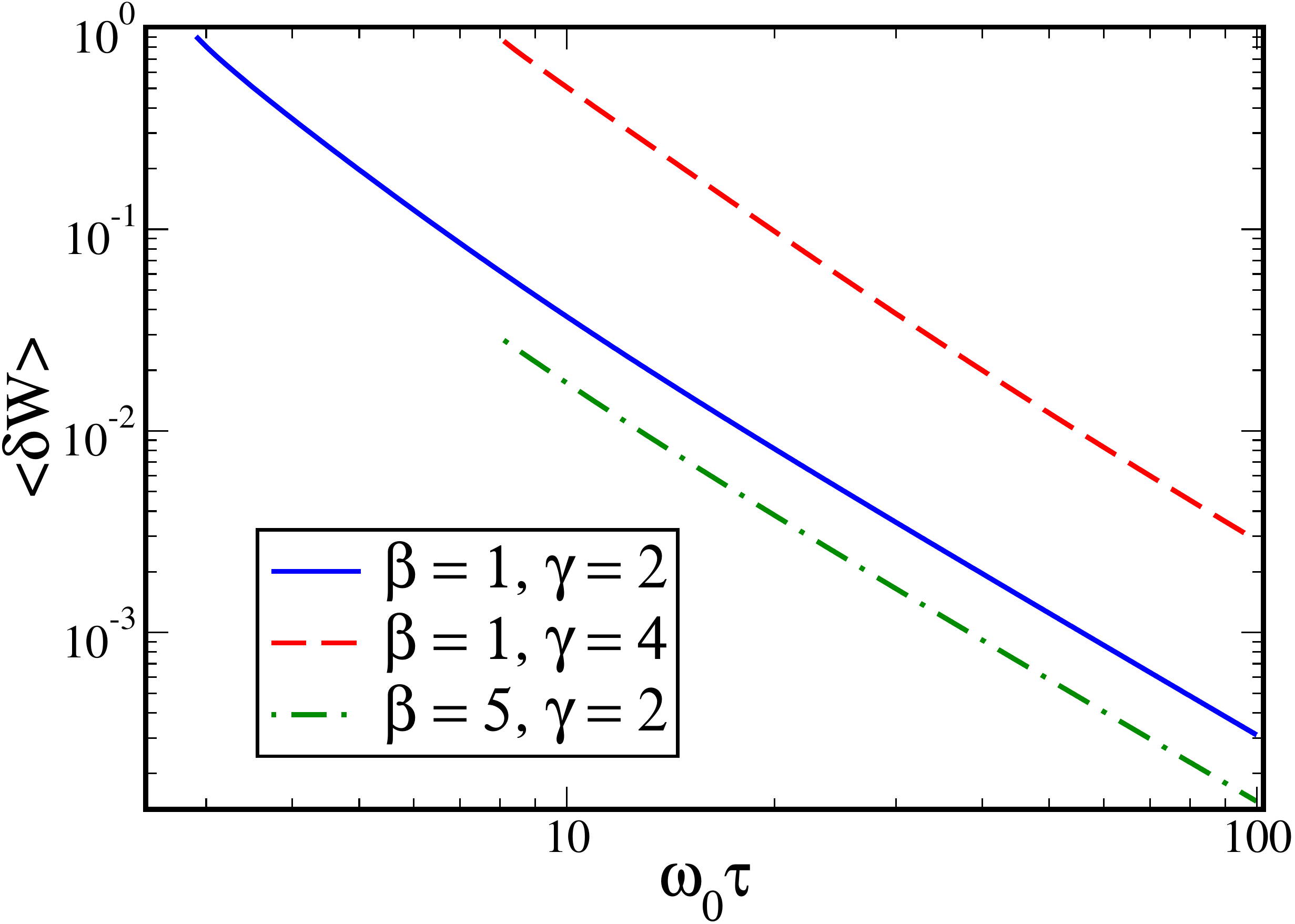, width=0.8\columnwidth}
\caption{{Quantum cost of running the super-adiabatic expansion stage of the quantum Otto cycle.} We plot the time-averaged deviation $\la\delta{W}\ra$ of the mean energy of the system from the adiabatic eigenenergies.
In all cases there is an effective power-law scaling of the form $\la\delta W\ra\sim1/\tau$. The cut-off time is such that the confining potential remains a trap along the process,
without the need for transiently inverting it to achieved the required speed up.
}
\label{dissW}
\end{figure} 
An explicit upper bound for the power of an engine run can be calculated using the fundamental limitations set by quantum speed limit, as shown in the Appendix.

\noindent
\section{Conclusions} We have demonstrated the possibility to perform a fully frictionless quantum cycle working in a finite-time only. Our proposal exploits the idea of shortcuts to adiabaticity, which allowed us to bypass the detrimental effects of friction on the compression and expansion stages in an important thermodynamical cycle such as the Otto cycle. We believe that our study embodies only {\it one} example of the potential brought about by the fascinating combination of shortcuts to adiabaticity and the framework for out-of-equilibrium dynamics of a quantum system. The possibilities to achieve maximum efficiency of a quantum engine with virtually no friction, yet within a finite operating time, is a tantalizing result with the potential to revolutionise the design of micro- and nano-scale motors operating at the verge of quantum mechanics. 

%
%
%
%
%
%


\acknowledgements

We greatly thanks  G. De Chiara, B. Damski, S. Deffner, R. Dorner, C. Jarzynski, E. Passemar and N. Sinitsyn for discussions and comments on the manuscript.
AdC is supported by the U.S. Department of Energy through the LANL/LDRD Program and a  LANL J. Robert Oppenheimer fellowship; 
JG acknowledges funding from IRCSET through a
Marie Curie International Mobility fellowship; MP thanks the UK EPSRC 
for a Career Acceleration Fellowship and a grant under the 
``New Directions for EPSRC Research Leaders" initiative (EP/G004579/1).

\section{Appendix}

\subsection*{Driving protocol of the super-adiabats} 

\begin{figure}[b]
\includegraphics[width=0.65\linewidth]{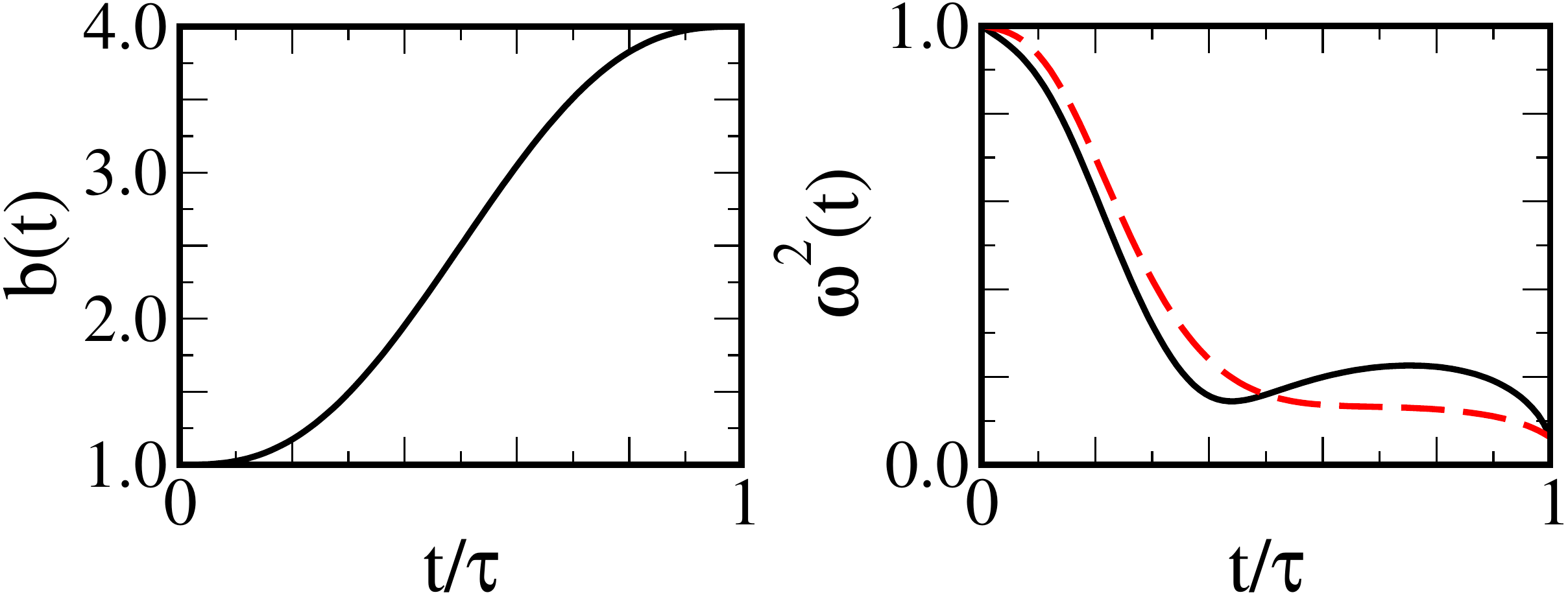}
\caption{
{Scaling factor and  driving frequency along a shortcut to an adiabatic expansion in a time-dependent harmonic trap.}
The scaling factor varies from $b(0)=1$ to $b(\tau)=4$ in a time scale $\tau=10/\om_0$ (solid line) and $20/\om_0$ (dashed line).
}
\label{b_om_t}
\end{figure}

In this part of the Appendix we illustrate the formal procedure for the determination of the scaling factor $b(t)$ used in the super-adiabatic steps of our proposal. The simplest interpolation of the actual solution of the Ermakov equation in the main Letter with the boundary conditions stated in the main text is found to be the polynomial 
\begin{equation}
\begin{aligned}
\label{b(t)}
b (t) =
6 \left(\gamma -1\right) s^5
-15 \left(\gamma-1\right) s^4 +10 \left(\gamma-1\right)s^3
+ 1
\end{aligned}
\end{equation}
with $s=t/\tau$. This solution guarantees that, in the adiabatic limit, $\dot{b}(t)\rightarrow 0$ and $b(t)\rightarrow b_{ad}$, while more generally the eigenstates of the initial oscillator will evolve according to the scaling law in Eq.~(6) of the main Letter. 
The modulation of $\om(t)$ is the responsible for the speed-up of the transformations performed along the super-adiabats. In turn, the implementation of such modulation is the price to pay for the achievement of such advantage. The explicit form $\om(t)$ can be extracted from the Ermakov equation, $\om^2(t)=\om_0^2/b^4(t)-\ddot{b}(t)/b(t)$, using Eq. (\ref{b(t)}) for $b(t)$, see Fig. (\ref{b_om_t}).
for sufficiently small values of $\om_0\tau$, $\om(t)$can become purely imaginary (requiring the inversion of the trap into an expelling barrier). 
The condition $\om^2(t)>0$ provides the cut-off time $\tau_c$ used in Fig.~3 of the main Letter. 

\subsection*{Nonequilibrium work fluctuations}

We next present the derivation of Eq.~(9) in the main Letter. Consider as reference states the fictitious equilibrium state $\rho_t^{\rm eq}$ and the adiabatic one $\rho_t^{\rm ad}$.
Then, in the adiabatic limit the average work can be rewritten as
\beqa
\la W_{\rm ad}(t)\ra&=&\tr[\rho_0(\hat{\mathcal{H}}(t)-\hat{\mathcal{H}}(0)]=\sum_np_n^0[\varepsilon_n(t)-\varepsilon_n(0)]\nonumber\\
&=&-\!\frac{1}{\beta}\sum_np_n^0\ln p_k^t\!+\frac{1}{\beta}\sum_n p_n^0\ln p_n^0\!-\frac{1}{\beta}\ln (Z_t/Z_0)\nonumber\\
&=&\frac{1}{\beta}S(\rho_{t}^{\rm ad}||\rho_{t}^{\rm eq})+\Delta F
\eeqa
with $Z_t=\tr[e^{-\beta \hat{\mathcal{H}}(t)}]$ the instantaneous partition function. For a general nonequilibrium process, the average work reads instead
\beqa
\la W\ra
&=&-\!\frac{1}{\beta}\sum_{nk}p_n^0p_{nk}^t\ln p_k^t\!+\frac{1}{\beta}\sum_n p_n^0\ln p_n^0\!-\frac{1}{\beta}\ln (Z_t/Z_0)\nonumber\\
&=&\!\frac{1}{\beta}S(\rho_{t}||\rho_{t}^{\rm eq})+\Delta F.
\eeqa
As a result, nonequilibrium deviations from the mean adiabatic work take the form
\beqa
\label{wirr1}
\delta W
=\frac{1}{\beta}[S(\rho_{t}||\rho_{t}^{\rm eq})-S(\rho_{t}^{\rm ad}||\rho_{t}^{\rm eq})].
\eeqa
It is worth considering an an alternative approach, where  $\rho_t^{\rm ad}$ is used as a reference state and the dynamics is restricted to the  class of 
self-similar processes \cite{Chen10,delcampo11,DB12}, for which conservation of the population in the mode $|n(t)\ra$ as a function of time $t$ is satisfied provided that
\beqa
\beta_t=\beta_0\varepsilon_n(0)/\varepsilon_n(t),
\eeqa
as it is the case for the adiabatic dynamics associated to the shortcuts discussed here. Under such condition the partition of the instantaneous 
equilibrium state remains constant $Z_t=Z_0=Z$.
Using $\rho_t^{\rm ad}$, the average work 
in the adiabatic limit reads
\beqa
\la W_{\rm ad}(t)\ra\!&=&\!\frac{1}{\beta_0}\sum_np_n^0\ln p_n^0\!-\!\frac{1}{\beta_t}\sum_{n}p_n^0\ln p_k^0\!-\!(\frac{1}{\beta_t}\!-\!\frac{1}{\beta_0})\ln Z\nonumber\\
&=&\frac{1}{\beta_{t}}S(\rho_{t})-\frac{1}{\beta_0}S(\rho_0)+\Delta F,
\eeqa
where we have introduced the von Neuman entropy $S(\rho)=-\tr[\rho\ln\rho]$ of an arbitrary state $\rho$. 
More generally, 
\beqa
\la W\ra\!&=&\!\frac{1}{\beta_0}\sum_np_n^0\ln p_n^0\!-\!\frac{1}{\beta_t}\sum_{k,n}p_{nk}^tp_n^0\ln p_k^0\!-\!(\frac{1}{\beta_t}\!-\!\frac{1}{\beta_0})\ln Z\nonumber\\
&=&\frac{1}{\beta_{t}}S(\rho_{t})-\frac{1}{\beta_0}S(\rho_0)+\frac{1}{\beta_{t}}S(\rho_{t}||\rho_{t}^{\rm ad})+\Delta F.
\eeqa
This leads to the following compact expression for nonequilibrium work deviations form the adiabatic path,
\beqa
\label{wirr2}
\delta W&=&\frac{1}{\beta_{t}}S(\rho_{t}||\rho_{t}^{\rm ad}).
\eeqa
The two expressions for $\delta W$, Eqs. (\ref{wirr1}) and (\ref{wirr2}), agree for self-similar processes and vanish at the end of the stroke (either 1 or 3 in Fig.~1 of the main Letter) both for a shortcut and  in the adiabatic limit. 

\subsection*{Upper bound to power through the quantum speed limit} 
The quantum speed limit for a driven quantum system  \cite{DL11}  allows us to derive an upper bound for the power of the engine. 
For simplicity, we can consider a equal-time shortcuts along  the two super-adiabats so that $\tau=\tau_1=\tau_3$. 
Then, it follows that 
\begin{equation}
{\cal P}\leq -\frac{\la W_{\rm ad, 1}(\tau)\ra+\la W_{\rm ad, 3}(\tau)\ra}{\hbar\,\mathcal{L}\left(\rho_\tau^{\rm eq},\rho_0\right)}{\rm max}\big\{E_\tau,\Delta E_\tau\big\}.
\end{equation}
where 
$E_{\tau}=\tau^{-1}\int_0^\tau d t\tr[\rho_t\hat{\mathcal{H}}(t)]$ with respect to the ground state energy,
$\Delta E_\tau=\tau^{-1}\int_0^\tau d t\,\{\tr[\rho_t \hat{\mathcal{H}}^2(t)]-\tr[\rho_t\hat{\mathcal{H}}(t)]^2\}^{1/2}$, and
 the angle in Hilbert space between initial and target states is
\begin{equation}
\mathcal{L}\left(\rho_0,\rho_\tau^{\rm eq}\right)=\arccos{\left( \sqrt{F\left(\rho_0,\rho_\tau^{\rm eq}\right)} \right) }
\end{equation}
in terms of the fidelity $F\left( \rho_0,\rho_\tau^{\rm eq}\right)=\left[{\sqrt{\sqrt{\rho_0}\,\rho_\tau^{\rm eq}\, \sqrt{\rho_0}}} \right]^2$.
In a super-adiabatic engine, $\la W\ra_{\rm ad, 1}+\la W\ra_{\rm ad, 3}$ equals
\be
\sum_{j=1,3}\la W_{\rm ad, j}(\tau)\ra=\frac{\hbar}{2}(\om_0-\om_\tau)
\big[\coth\frac{\beta_c\hbar\om(\tau)}{2}-\coth\frac{\beta\hbar\om_0}{2}\big],
\ee
where $\beta_c$ is the inverse temperature of the cold bath during stage $2$.

\end{document}